\documentstyle[preprint,aps,epsfig,tighten]{revtex}
\begin{document}
\title{Extrapolation of the astrophysical $S$ factor for 
       $^7$Be$(p,\gamma)^8$B to solar energies} 
\author{B.K.~Jennings, S.~Karataglidis, and T.D.~Shoppa}
\address{TRIUMF, 4004 Wesbrook Mall, Vancouver, British Columbia, Canada, 
V6T 2A3}
\date{\today}
\maketitle
\begin{abstract}
We investigate the energy dependence of the astrophysical $S$ factor
for the reaction $^7$Be$(p,\gamma)^8$B, the primary source of
high-energy solar neutrinos in the solar $pp$ chain. Using simple
models we explore the model dependence in the extrapolation of the
experimental data to the region of astrophysical interest near
20~keV. We find that below approximately 400~keV the energy dependence
is very well understood and constrained by the data for the elastic
scattering of low energy neutrons from $^7$Li. Above 400~keV nuclear
distortion of the wave function of the incident proton introduces a
significant model dependence. This is particularly important for the
$s$-wave contribution to the $S$ factor. The extracted value of $S(0)$
is $19.0 \pm 1.0 \pm 0.2$~eVb. The first error is experimental while
the second is an estimate of the theoretical error in the
extrapolation.
\end{abstract}

\section{Introduction}
\label{introduction}

The $^7$Be($p,\gamma$)$^8$B reaction, at energies of approximately
20~keV, plays an important role in the production of solar
neutrinos\cite{bahcall}. The subsequent decay of the $^8$B is the
source of the high energy neutrinos to which many solar neutrino
detectors are sensitive. The cross section for this reaction is
conventionally expressed in terms of the $S$ factor which is defined
in terms of the cross section, $\sigma$, by:
\begin{equation}
S(E) = \sigma(E) E \exp{\left[ 2\pi\eta(E) \right]}\;,
\label{eq-define}
\end{equation}
where $\eta(E) = Z_1 Z_2 \alpha \sqrt{\mu c^2/2E}$ is the Sommerfeld
parameter, $\alpha$ is the fine structure constant, and $\mu$ is the
reduced mass. The definition of the $S$~factor eliminates from it most
of the energy dependence due to Coulomb repulsion by factoring out the
penetration to the origin of a particle in the Coulomb potential of a
point charge.  However, it does not make the $S$~factor energy
independent, as there are still energy dependences due to the
structure of the final bound state, resonances and the attenuation of
the barrier by the nuclear mean field.  The reaction rate, obtained by
folding the thermal distribution of nuclei in the stellar core with
the cross section, peaks at approximately 20~keV. Because the cross
section diminishes exponentially at low energies, the only method of
obtaining information about the $S$ factor at energies of
astrophysical interest is to extrapolate data taken at experimentally
accessible energies ($E > 100$~keV). To do the extrapolation reliably
we must understand the physics associated with the $S$ factor.

\begin{figure}
\centering\epsfig{file=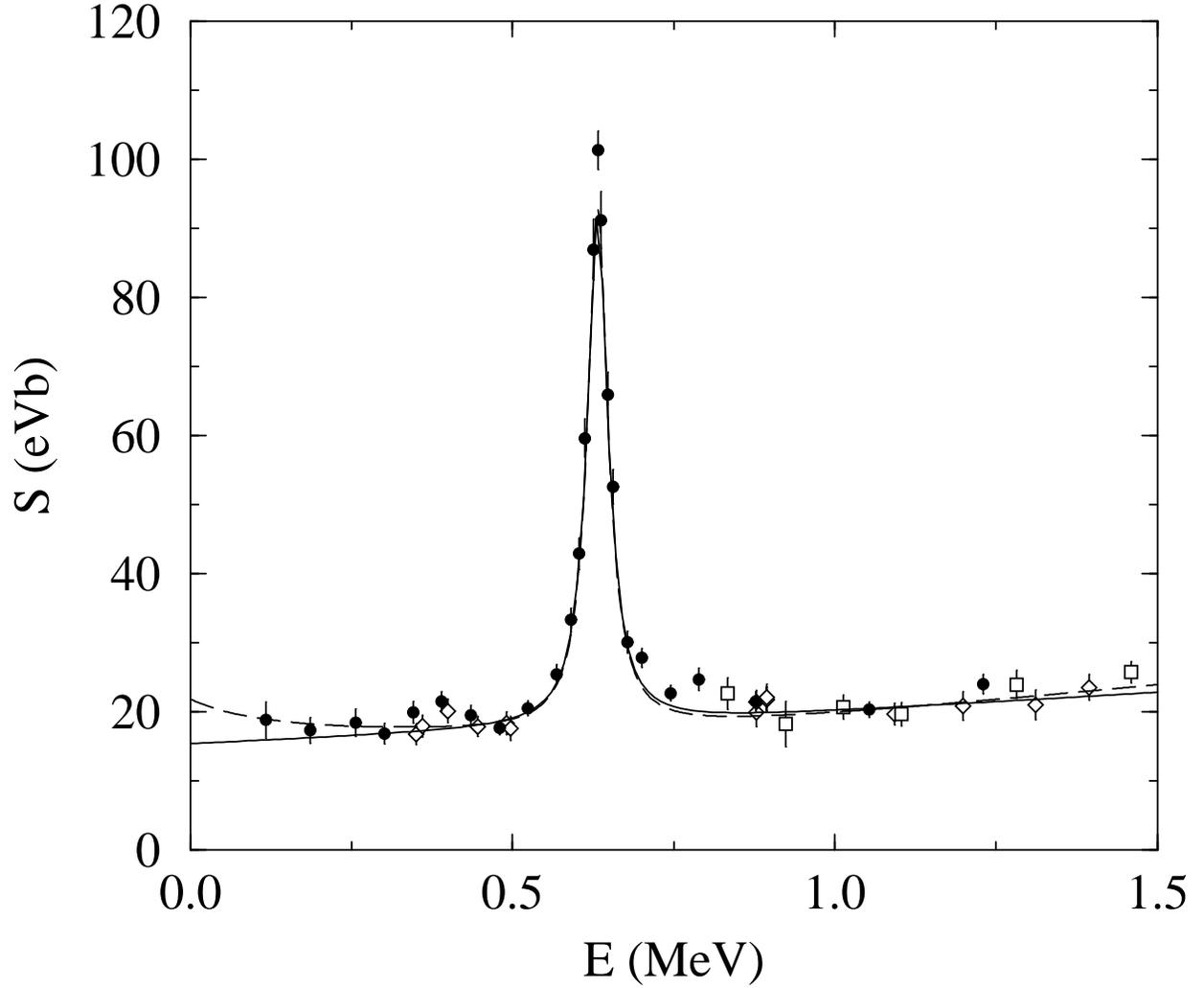,width=\linewidth}
\caption[]{Different fits to the experimental $S$-factor data. The
solid curve is a straight line plus the resonance while the dashed
curve is a calculation with a hard sphere cut-off radius of 4.1 fm
plus the resonance. The $\chi^2 = 0.9$ in both cases. The data are from
Vaughn {\em et al.} \protect\cite{Vaughn} (squares),
Filippone {\em et al.} \protect\cite{Filippone} (circles) and
Hammache {\em et al.} \protect\cite{Hammache} (diamonds). }
\label{fig-fits}
\end{figure}

To illustrate the problem of extrapolating the data to astrophysical
energies we show in Fig.~\ref{fig-fits} a fit to the experimental data
\cite{Filippone,Vaughn,Hammache} that uses just a straight line and a
fit with a calculation that includes $s$-wave nuclear distortion
through a hard sphere potential of radius 4.1~fm. The straight line
and potential model fits are displayed by a solid and dashed line
respectively. The latter calculation will be described in more detail
in Sec.~\ref{sec-pades}. In both cases a Breit-Wigner resonance is
included. As can be seen from the figure both fits to the data are
equally good; $\chi^2 = 0.9$ in each case. However, there is a marked
difference in the $S$ factors at 20~keV: 15.3 and 21.0~eVb for the
straight line fit and the hard sphere model, respectively; a 37\%
variation in the extrapolated value. Such a large difference must be
understood if reliable extrapolations are to be made and a variety of,
sometimes conflicting, models
\cite{Christy,Calvin,barker1,barker2,barker3,tombrello,robertson,kim,desc,nunes,jennings}
have been developed for this purpose.

The dashed curve in Fig.~\ref{fig-fits} shows an upturn in the $S$
factor at threshold. This is a feature common to all of the model
calculations of the astrophysical $S$ factor for the
$^7$Be($p,\gamma$)$^8$B reaction (with the exception of the straight
line fit). It has been established \cite{jennings} that this behavior
stems from a pole in the $S$ factor when the photon energy,
$E_{\gamma}$ vanishes.

We develop two key concepts for the description of the astrophysical
$S$ factor: the pole where the photon energy vanishes and an effective
hard sphere radius. These concepts will be developed and explored by
modeling the complicated multi-dimensional many-body system with
simple one-body models.  The pole term describes and is dominated by
Coulomb physics. It depends on nuclear physics through the separation
energy, the asymptotic normalization of the final state wave function
and the spectroscopic factors. The separation energy of the valence
proton from $^8$B is 137.5~keV \cite{Aj88} and determines the pole
location. The asymptotic normalization and the spectroscopic factor
combine in an asymptotic strength parameter, defined in
Sec.~\ref{sec-pole}, to give the residue of the pole. This will be
determined by a fit to the $S$-factor data. An effective hard sphere
repulsion is introduced to approximate how the nuclear physics
influences the energy dependence, and is related primarily to the
non-resonant phase shift of the initial scattering state. The
effective hard-sphere radius is determined by comparing the
hard-sphere model to more complete models such as potential models.
Elastic scattering data from protons on $^7$Be, if it existed, would
help constrain the potential models and through them the effective
hard shpere radius.  As such data are presently unavailable, the
constraints must come from the elastic scattering of neutrons from
$^7$Li, the mirror system. The concepts of the pole and the hard
sphere repulsion lead to a simple rational approximation for the
energy dependence of the $S$ factor, which encompasses the dominant
physics at low energy.

This simple approach breaks down as the energy increases and the
capture becomes more sensitive to the internal structure of the $^7$Be
core.  We explore the range of validity of the simple one-dimensional
model by comparing it with more sophisticated potential, $R$--matrix,
and microscopic cluster models. In general, the simple approach agrees
with the potential and $R$--matrix models over a larger energy range.
Alternatively, within its range of validity, we may use the hard
sphere model based on an effective hard sphere radius and Coulomb
physics to understand and critique other models.

We present a brief review of the formalism in Sec.~\ref{sec-pole} and
show how the pole in the $S$ factor arises. In Sec.~\ref{sec-pades}
the hard sphere model is presented and used to derive a simple
expression for the energy dependence of the $S$ factor.  The hard
sphere model is compared to the cluster model calculations in
Sec.~\ref{sec-models}. In Sec.~\ref{sec-data} we use the final
arbiter, the experimental data, to discriminate between the models and
determine which is best for the extrapolation.  In
Sec.~\ref{sec-conclusions} we present our best estimates for the $S$
factor near threshold and draw some conclusions.

\section{The pole term}
\label{sec-pole}

Most calculations of the $S$ factor follow the pioneering work of
Christy and Duck \cite{Christy} to which we refer the reader for more
details. Here we present a brief overview of the model. The $E1$
contribution to the $S$ factor for the $^7$Be($p,\gamma$)$^8$B
reaction may be written as
\begin{equation}
S = C (I_0^2 + 2 I_2^2) E_\gamma^3 \left( J_{11} \beta_{11}^2 + 
  J_{12} \beta_{12}^2 \right) \frac{1}{1-e^{-2 \pi \eta}} \;,
\label{eq-one}
\end{equation}
where
\begin{eqnarray}
I_l & = & \int^\infty_0 r^2 dr \; r \; \psi_{il}(r) \psi_f(r) \label{eq-int}\\
C   & = & \frac{5\pi}{9}  \frac{1}{(\hbar c)^3} (2 \pi \eta k) e^2 \mu^2
 \left( \frac{Z_1}{M_1} - \frac{Z_2}{M_2} \right)^2 \;.
\end{eqnarray}
In Eq.~(\ref{eq-one}), $J_{LS}$ is the spectroscopic factor for a
given angular momentum, $L$, and channel spin, $S$, $\beta_{LS}$ is
the asymptotic normalization of the bound state wave function,
$E_\gamma$ is the photon energy, and $k$ is the momentum of the
incident proton.  The extra factor of $r$ in the integrand comes from
the photon wave function. The final bound state wave function
$\psi_{f}(r)$ is normalized in the asymptotic region to $\psi_{f}(r) =
W_{\alpha,l}(\kappa r)/r$ while the initial wave function reduces to
the regular Coulomb wave function divided by $kr \sqrt{2 \pi
\eta}/(e^{2 \pi \eta}-1)$. The unusual choice of normalizations is
just to simplify the mathematics and generate integrals that are
well-behaved at threshold.  The initial state has both Coulomb and
nuclear distortions. The Coulomb distortions are large and give the
penetration factor included in the definition of the $S$ factor,
Eq.~(\ref{eq-define}). They are included in all calculations. The
nuclear distortions are much smaller but they are important and
introduce a significant model dependence into the calculations, as
described in the next section.

The absolute magnitude of the $S$ factor is determined primarily by
the spectroscopic factor and the asymptotic normalization (see also
Ref.~\cite{Xu}). The spectroscopic factor contains many-body aspects
of the problem and is calculable from standard shell model theory. The
asymptotic normalization also depends on the many-body wave function,
but is far more difficult to estimate from first principles: it
requires detailed knowledge of how the 8-body wave function extends
beyond the nuclear potential and its mapping to the Whittaker function
in this region.  This may be estimated crudely by approximating that
behavior by using a suitably chosen Woods-Saxon wave function for the
weakly bound proton. Instead we treat the overall factor,
$A_n=J_{11}\beta_{11}^2 + J_{12}\beta_{12}^2$, as a free parameter,
which is independent of energy, and determined by the $S$-factor
data. For simplicity we will refer to this combination of asymptotic
normalization and spectroscopic factor as the asymptotic strength.

To investigate the behavior of the integrals in Eq.~(\ref{eq-one}), we
first consider $\psi_{f}(r) = W_{\alpha,l}(\kappa r)/r$ for all radii
and take $\psi_{i0}(r) = F_0(k r)/\{ kr \sqrt{2 \pi \eta} / (e^{2 \pi
\eta}-1) \}$. The $s$-wave integral then becomes
\begin{equation}
I_0 = \int^{\infty}_0 dr \; r \frac{ W_{\alpha,l}(kr) F_0(kr) }{k
\sqrt{2 \pi \eta} } \left( e^{2\pi\eta} - 1 \right) \; .
\label{eq-coul}
\end{equation}
The integral is smooth as $k$ passes through zero and diverges as $k
\rightarrow i\kappa$ ($E \rightarrow -E_{\rm B}$). The nature of the
divergence is determined by the asymptotic forms of the Coulomb wave
function and Whittaker function for large $r$. There the Whittaker
function is proportional to $r^{- |\eta k|/\kappa} e^{-\kappa
r}$\cite{Christy} ($\eta k$ is independent of $k$). Above threshold
the Coulomb wave function oscillates at large radii, however below
threshold it is exponentially growing and is proportional to
$r^{|\eta|} e^{|k| r}$. Thus the behavior of the integrand at large
radius is
\begin{equation} 
       r^{1-|\eta k|(1/\kappa-1/|k|)} \exp[-(\kappa-|k|)r] 
\end{equation}
and the integral diverges as 
\begin{equation}
I_0 \sim 1/(\kappa-|k|)^2 \sim 1/(E_{\rm B}+E)^2 = 1/E_\gamma^2\;.
\end{equation}
The $S$ factor is proportional to $I_0^2E_{\gamma}^3$, and gives rise
to a simple pole in $S$ at $E_{\gamma} = 0$. However, the first
correction term is not simply $1/E_\gamma$ but rather of the form
$\left( 1 + c \log{E_\gamma} \right)/E_{\gamma}$, the logarithmic term
coming from the $r^{-|\eta k|(1/\kappa-1/|k|)}$ factor. Both the
leading and first correction terms are determined purely by the
asymptotic behavior of the wave functions. The second correction term,
of order $E_\gamma^0$, is not determined purely by the asymptotic
value of wave function alone but also depends on the wave function at
finite $r$.
 
The presence of the pole suggests the $S$ factor may be parametrized
as a Laurent series:
\begin{equation}
S = d_{-1} E_\gamma^{-1} + d_0 + d_1 E_\gamma +\ldots
\label{eq-laurent}
\end{equation} 
The coefficients of the first two terms, $d_{-1}$ and $d_0$, are
determined purely by the asymptotic forms of the wave functions while
the third coefficient, $d_1$, is also dependent on the short range
properties of the wave functions. The validity of such an
approximation is discussed below.

\section{One-body models and rational approximations}
\label{sec-pades}

\begin{figure}
\centering\epsfig{file=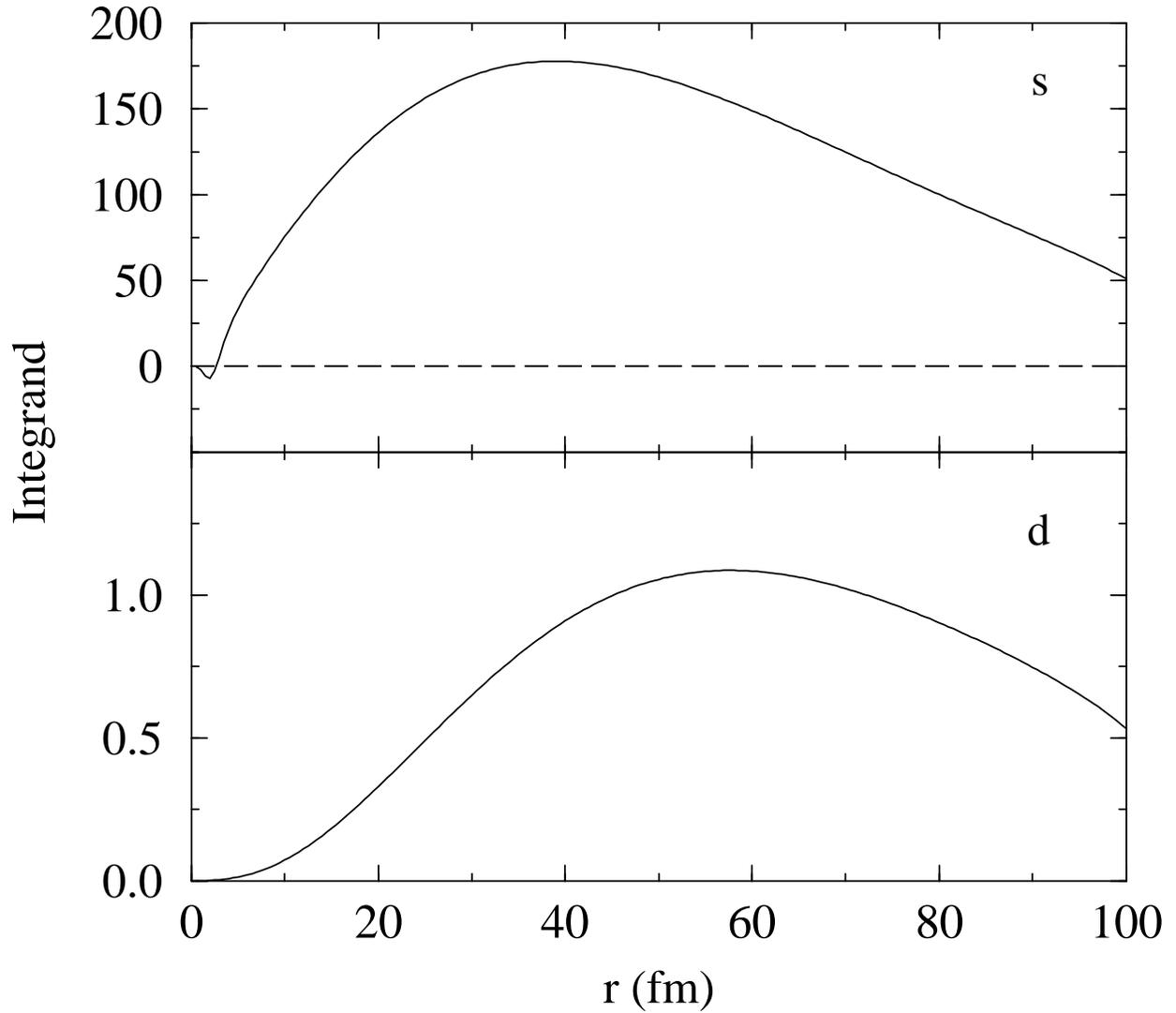,width=\linewidth}
\caption[]{The integrand for the $s$-wave and $d$-wave contributions to the
$S$ factor at 0.0~MeV. The calculations were done with a Wood-Saxon
potential.}
\label{fig-0p5}
\end{figure}

The energy dependence in the $S$ factor enters through the $s$- and
$d$-wave integrals, $I_0$ and $I_2$, and the phase-space factor,
$E_{\gamma}^3$. To investigate that behavior, we present in
Fig.~\ref{fig-0p5} the integrands, Eq.~(\ref{eq-int}), for both the
$s$ and $d$ waves. A Woods-Saxon potential model, denoted B1, whose
radius (2.39~fm) and diffuseness (0.65~fm) were taken from Barker
\cite{barker1}, was used to calculate the bound state and nuclear
distortions. A potential depth of $-$46.6~MeV was chosen to reproduce
the binding energy of the final state. No spin-orbit force was
included. The integrands are peaked at very large radii: 40~fm and
55~fm for the $s$- and $d$-wave integrands, respectively, and extend
well beyond 100~fm. As this is well outside the range of the nuclear
potential ($r_{rms} = 2.48 \pm 0.03$~fm for $^7$Be \cite{Ta85}), the
capture is purely Coulombic. To ensure complete convergence in our
calculations we integrated to 1000~fm. The small negative contribution
and the node in the $s$-wave integral near 0~fm arise from the effects
of the nuclear distortion. The distortion in the $d$-wave component is
negligible.

\begin{figure}
\centering\epsfig{file=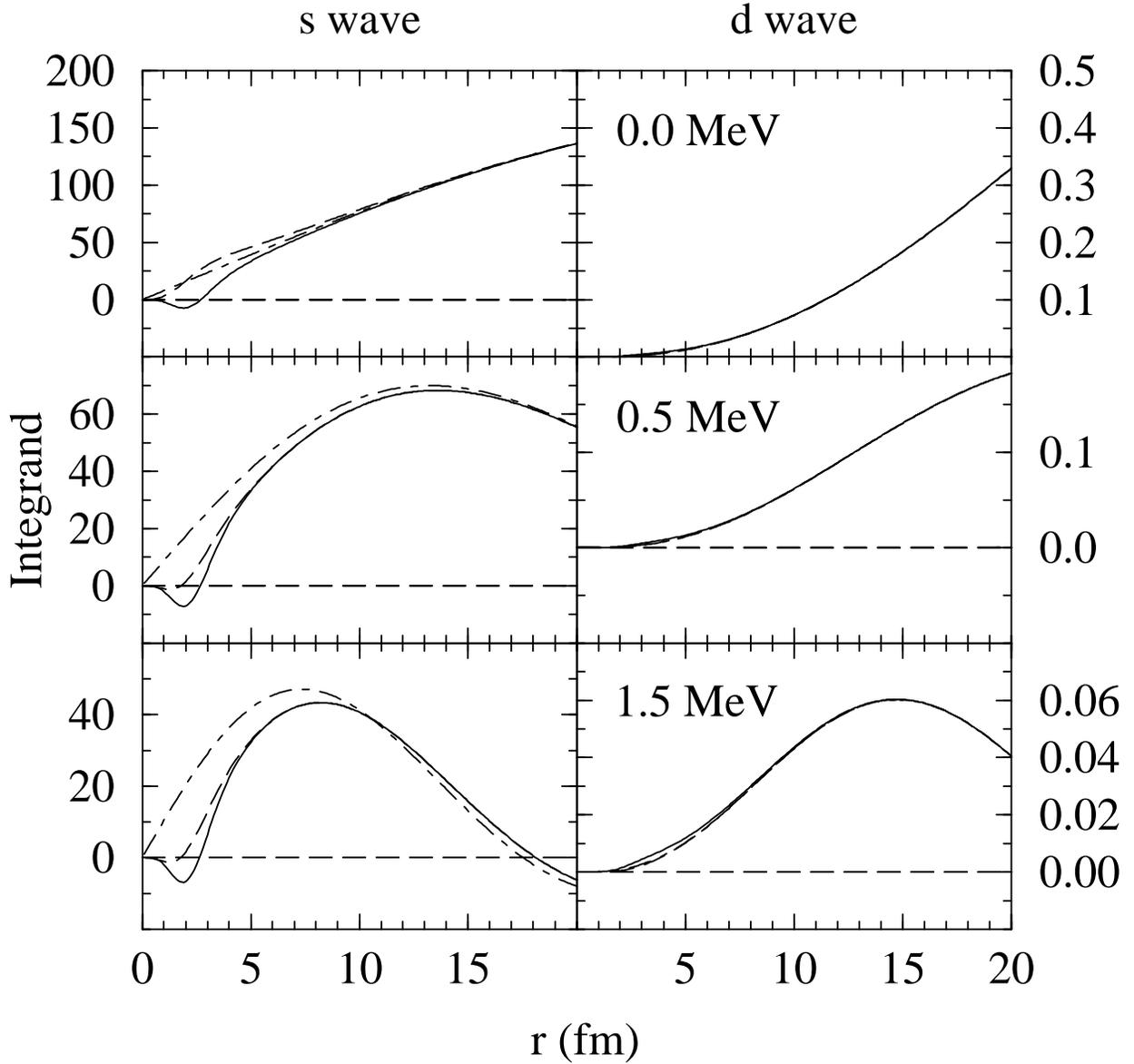,width=\linewidth}\vspace{0.5cm}
\caption[]{The integrand for the $s$- and $d$-wave contributions to the
$S$ factor at 0.0~MeV, 0.5~MeV, and 1.5~MeV. The solid line is a full
calculation with a Woods-Saxon potential. The dot-dashed line has no
nuclear distortion of the incoming wave and uses a Whittaker function
for the bound state. The dashed curve is the extrapolation of full
calculation to short distance using only the Coulomb potential.}
\label{fig-1p5}
\end{figure}

We show the integrands for a range of energies in
Fig.~\ref{fig-1p5}. The peaks in the integrands, which are displayed
by the solid lines, are outside the range of the nuclear force even at
1.5~MeV. Two other calculations, in which the nuclear distortion is
varied, are also presented.  The pure Coulomb calculation,
Eq.~(\ref{eq-coul}), contains no nuclear distortions and is displayed
by the dot-dashed line. The other curve (dashed line) matches the full
calculation at large distances so it has the same nuclear phase shift
at large radii but the initial wave function is integrated to small
radii using just the Coulomb potential. In the $s$-wave integrands the
nuclear distortions play an important role, especially near the
origin. These distortions produce a node in the scattering wave
function and give a repulsive phase shift. The node is necessary to
make the scattering wave function orthogonal to the bound $0s$-shell
protons in the $^7$Be core. That orthogonality is preserved only if
the node in the scattering state is at a radius where the bound state
wave function is still appreciable. Hence the node will be close to
but inside the nuclear radius. There are no bound $0d$-shell protons
in $^7$Be and hence no node in the $d$-wave integrand. Consequently
the $d$-wave phase shift is small and attractive.

The integrals corresponding to the integrands plotted in
Fig.~\ref{fig-1p5} are given in Table~\ref{table-3}. The largest
change in the $s$-wave integral comes from the nuclear phase shift at
large radii, as is evident when partial nuclear distortion is
introduced. The additional distortion coming from the short-range
nuclear potential produces a smaller but still significant change in
the integral. This is common to the integrals evaluated at 0.5 and
1.5~MeV. In the case of the $d$-wave integral, the total effect of
nuclear distortions is quite small, at most 2\%, even at 1.5~MeV.

Since the $S$ factor is sensitive to the phase shift the potential
should reproduce the nuclear phase shifts. Unfortunately, there are no
experimental data for the scattering of protons from $^7$Be from which
the phase shifts may be determined. However, data are available for
the scattering in the mirror system, $^7$Li--$n$, for which we follow
Barker's analysis \cite{barker1} to determine the potential
depths. From the elastic scattering of thermal neutrons from $^7$Li,
the scattering lengths are $a_1 = 0.87 \pm 0.07$~fm and $a_2 = -3.63
\pm 0.05$~fm \cite{Ko83}, where $a_S$ is the scattering length for the
channel spin $S$. The depths of the potentials are adjusted to fit
these scattering lengths, giving $-46.58$ MeV and $-56.21$ MeV for the
$S = 1$ and 2 potentials, respectively. The $S = 1$ potential depth is
very similar to the one we have used for the first Woods-Saxon
potential model but the $S = 2$ potential is significantly
stronger. These nuclear potentials are then used in the calculation of
the $S$ factor for $^7$Be($p,\gamma$). This assumes isospin symmetry
for the nuclear mean field.  The contributions for the two channel
spins are combined using Barker's spectroscopic factors
\cite{barker1}. We refer to this potential model as B2.

In Fig.~\ref{fig-1p5}, we see that the $s$-wave nuclear distortion is
dominated by the node in the scattering state wave function. This
suggests that we may construct a simple model, the hard sphere model
discussed in the introduction, where the initial state wave function
is zero inside some radius, $r_c$, and a pure Coulomb wave outside. We
impose the boundary condition that the wave function be zero at
$r_c$. This generates a phase shift and is equivalent to having an
infinitely repulsive potential with a radius $r_c$. The $d$-wave
scattering state is taken to be an undistorted Coulomb wave
function. The bound state is assumed to be a pure Coulomb state,
described by a Whittaker function, for all radii.

We saw for the potential model that it was necessary to have different
potential depths for different channel spins. For the hard-sphere
model this would suggest that we use different cut-off radii for
different channel spins. However the simple model does not justify
such elaborations and we find that with just a single, suitably
chosen, radius we can reproduce the low energy results from a given
potential model. The use of a single radius does, however, weaken the
connection between the hard-sphere radius and the elastic scattering
phase shift. We make the connection only through the intermediary of
the potential model.

\begin{figure}
\centering\epsfig{file=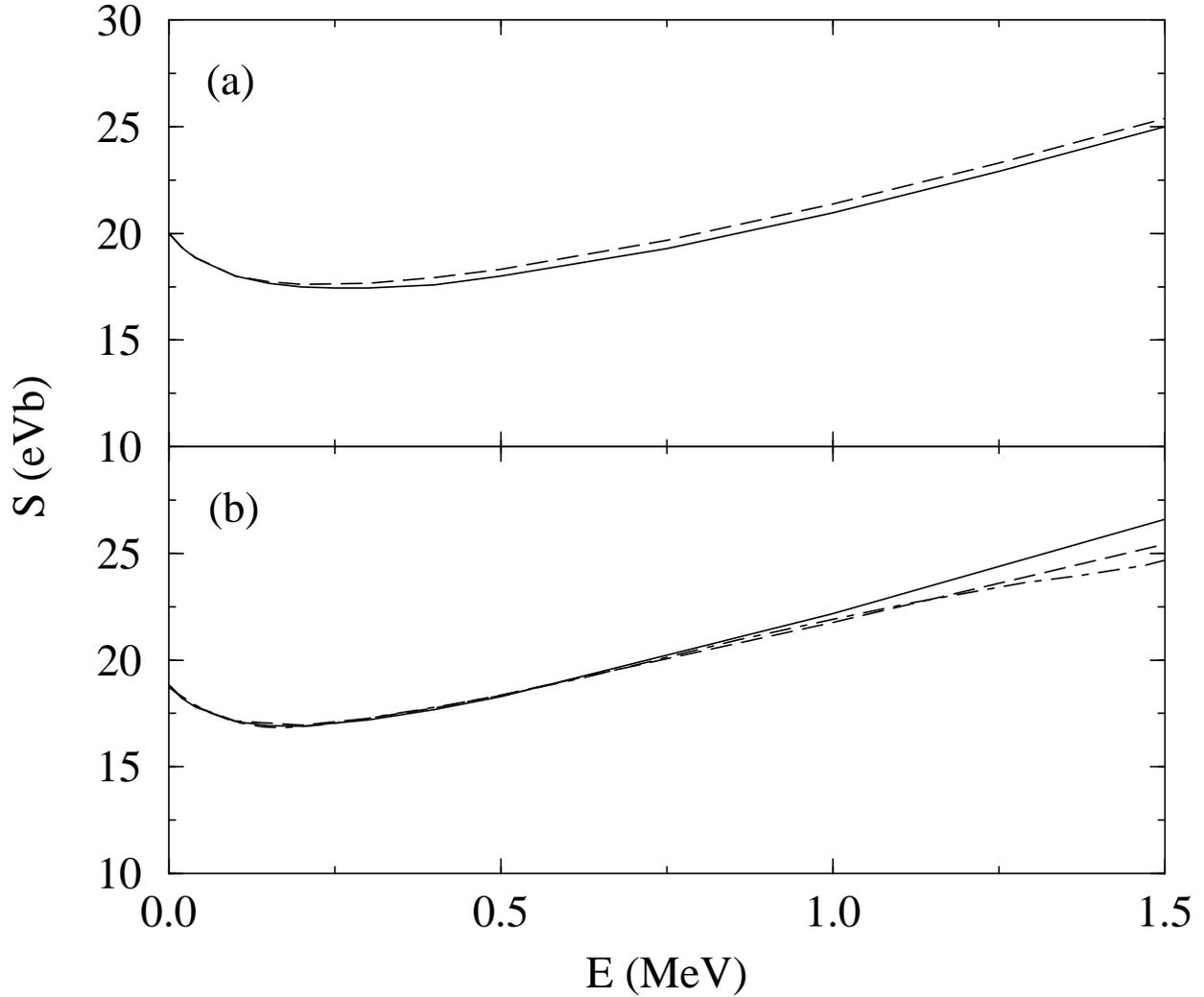,width=\linewidth}
\caption[]{The $S$ factor as a function of energy for different one-body
models. All curves are normalized to the same value at 0.1~MeV. In (a)
the solid line is a hard sphere model with $r_c = 2.4$~fm while the
dashed line is the B1 Woods-Saxon model. In (b), the model with $r_c =
1.0$~fm, the B2 model and the $R$-matrix model of Barker are displayed
by the solid, dashed, and dot-dashed line, respectively.}
\label{fig-two}
\end{figure}

The $S$ factors from the hard-sphere model are compared to the $S$
factor of other one-body models in Fig.~\ref{fig-two}. In
Fig.~\ref{fig-two}(a), the solid curve is the result of the hard
sphere model, with $r_c = 2.4$~fm while the dashed curve displays the
result of the Woods-Saxon potential, B1. The curves have been
normalized to agree at $E = 100$~keV. There is remarkable agreement
between the two results up to 1.5~MeV suggesting that the hard sphere
model encapsulates the physics of the Woods-Saxon potential model. In
Fig.~\ref{fig-two}(b) we show the results of a hard sphere model
calculation with radius, $r_c=1.0$~fm, the calculation with the B2
potential, and Barker's $R$-matrix calculation with $a = 4$~fm
\cite{barker3}. That choice of $a$ is predicated on the result that,
in the $R$-matrix formalism, the matching radius should be roughly the
sum of the radius and diffuseness of the potential \cite{Vo96}. The
level of agreement is again quite good.

One advantage of the hard sphere model is that it is possible to do an
explicit Taylor series expansion about $E=0$~MeV and obtain directly
the derivatives of $S$ at threshold. Following Williams and
Koonin\cite{williams} we employ the Bessel function
expansion\cite{abm} of the Coulomb wave functions to generate the
Taylor series expansion. Each term involves $E$-independent radial
integrals of Bessel functions, powers, and Whittaker functions. For a
hard sphere radius of 2.4~fm the integrals may be done to yield the
series:
\begin{equation}
S(E)/S(0) = 1 - 1.917 E + 15.69 E^2 - 110.28 E^3 + 774.1 E^4 +
\ldots \;
\label{eq-taylor} 
\end{equation}
where $E$ is in MeV. The coefficients are increasing in size and
alternate in sign. Given the pole in the $S$ factor, the radius of
convergence is $E=E_B=137.5$~keV.  We stress that the coefficients in
Eq.~(\ref{eq-taylor}) were not obtained by a fit of the $S$ factor
over a finite energy region but rather through an explicit series
expansion of the $S$ factor in powers of the energy.

A similar expansion has been used by Baye {\em et al.}\cite{descb} to
obtain the first derivative of the $S$ factor at threshold.  They
utilize a slightly different Bessel function expansion for the Coulomb
functions. It can be obtained from that used by Williams and
Koonin\cite{williams} by using the recurrence relations for the Bessel
functions.  For the choice of cutoff radius of Baye {\em et al.},
2.0~fm, we reproduce their numerical results.

A more convergent and pedagogically useful expansion may be
developed. Motivated by the Laurent series of Eq.~(\ref{eq-laurent}),
the Pad\'e approximant discussion of Ref.~\cite{jennings} and the
knowledge that there is a pole at $E_\gamma=0$~MeV, we Taylor series
expand $E_\gamma S$. This removes the effect of the simple pole in the
expansion. To recover $S$ we divide by $E_\gamma$ thus obtaining a
rational approximation:
\begin{eqnarray}
S(E)/S(0)& = & \frac{0.1375+0.7361 E+0.2392 E^2 +\ldots}{0.1375+E}
   \label{eq-padez}\\
	 & \approx & \frac{a}{E_B + E} + b + cE \label{eq-pade},
\end{eqnarray}
with $a = 0.0408$~MeV, $b = 0.7033$, and $c = 0.2392$~MeV$^{-1}$. By
construction, $a/E_B + b = 1$.  The rational approximation,
Eqs.~(\ref{eq-padez}) or (\ref{eq-pade}), is very similar to a Pad\'e
approximant.  The Pad\'e approximant is a ratio of polynomials with
all the parameters determined by fitting the derivatives at the
expansion point. However in Eq.~(\ref{eq-padez}), the position of the
pole is fixed by the binding energy. As a result, for the same order
polynomials one less derivative is required.  The coefficients in the
rational approximation, Eq.~(\ref{eq-padez}), are growing much more
slowly than in the Taylor series expansion, Eq.~(\ref{eq-taylor}).
This is due to the better convergence of the rational
approximation. The coefficients do, however, begin to grow more
rapidly after the cubic term in the numerator.

The accuracy of the rational approximation, [Eq.~(\ref{eq-pade})], is
shown in Table~\ref{table-pade}. This approximation is valid to better
than 1\% up to 400~keV. By comparison, the Taylor-series expansion and
the logarithmic derivative expansion break down below 100~keV as
expected given the radius of convergence.  Although all approaches are
accurate in the astrophysical region near 20~keV only the rational
approximation is accurate out to the region which is accessible by
experiment.

The coefficients, $a$, $b$ and $c$, in Eq.~(\ref{eq-pade}) are given
in Table~\ref{table-padeb} for a variety of models. In the case of the
hard sphere models, the pole and constant term are the same to within
0.5\% while the linear term varies by a factor of almost six. This
confirms that the constant and pole term are coupled in an almost
model independent manner while the linear term is strongly dependent
on initial-state nuclear distortions.

The first and second logarithmic derivatives are also given in
Table~\ref{table-padeb} for comparison with the results of Williams and
Koonin\cite{williams}.  They have a binding energy of 136~keV and use
a hard sphere model with $r_c=4.1$~fm in both the $s$ and $d$
waves. We agree with Barker\cite{barker2} that this choice of radius
is poorly motivated and we find that none of the Woods-Saxon or
generator coordinate models considered herein are consistent with a
hard sphere model with $r_c > 3$~fm. Williams and Koonin are also
missing a factor of 2 for the $d$-wave term in their Eq.~(1).

Also in Table~\ref{table-padeb}, we make the comparison of the
parameterization of the hard sphere models to those for three
different Woods-Saxon calculations. The parameterization,
Eq.~(\ref{eq-pade}), for the Woods-Saxon models was determined from
fitting the $S$ factor at 0, 20, and 40~keV. For two of the
calculations, we use the B1 and B2 models introduced previously. For
the third, the radius and diffuseness parameters were obtained from
Tombrello \cite{tombrello} while the potential depth was adjusted to
reproduce the binding energy of the final state. No spin-orbit force
was included.  The third calculation is denoted as T. The results for
the B1 and T models are quite close to those of the hard sphere model
with $r_c = 2.4$~fm.  While the $a$ and $b$ coefficients in the B2
model are consistent with those of the other Woods-Saxon models, the
linear term, $c$ is closer to the hard sphere model with $r_c =
1.0$~fm. This is consistent with the agreement we have seen in
Fig.~\ref{fig-two} between the B2 model and the hard sphere model with
that radius.

Attempts have been made to obtain the derivatives at threshold by a
quadratic fit to either $S$ or $\log{S}$ over an extended energy
range. The derivatives obtained by this method tend to disagree among
themselves and with our results. The derivatives from two such fits
are shown in the last two rows of Table~\ref{table-padeb}. The first
is a fit to $\log{S}$ by Barker\cite{barker2} over the energy range 0
to 100~keV. He uses a Woods-Saxon potential model to obtain $S$. The
second is from Adelberger {\em et al.} \cite{adelberger}. There $S$ is
obtained from a generator coordinate calculation \cite{Calvin} and fit
over the energy range 20 to 300~keV. As shown in
Table~\ref{table-pade}, Taylor series expansions about the origin are
not valid over the energy ranges used for the fits. We find that in
order to accurately determine both the first and second derivatives at
the origin it is necessary to restrict the fit region to less than
10~keV.  The Taylor series expansion converges well in this energy
region.  We can, however, qualitatively reproduce the numbers of
Barker and Adelberger {\em et al.} for the derivatives from our models
if we use their fit regions. Thus the differences in the numbers
obtained are not primarily from differences in the models but rather
due to how the derivatives were obtained. They indicate the
sensitivity to the fit range chosen.

While a quadratic form does not work well near threshold it is quite
good if that region is excluded. For example, a fit to the $S$ factor
over the range 30 to 300~keV is accurate to better than 0.4\% except
very near the end points. Using this fit to extrapolate to threshold
gives almost a 3\% error in $S(0)$; not too surprising given that the
quadratic form ignores the existence of the pole at $E = -137.5$~keV.

\begin{figure}
\centering\epsfig{file=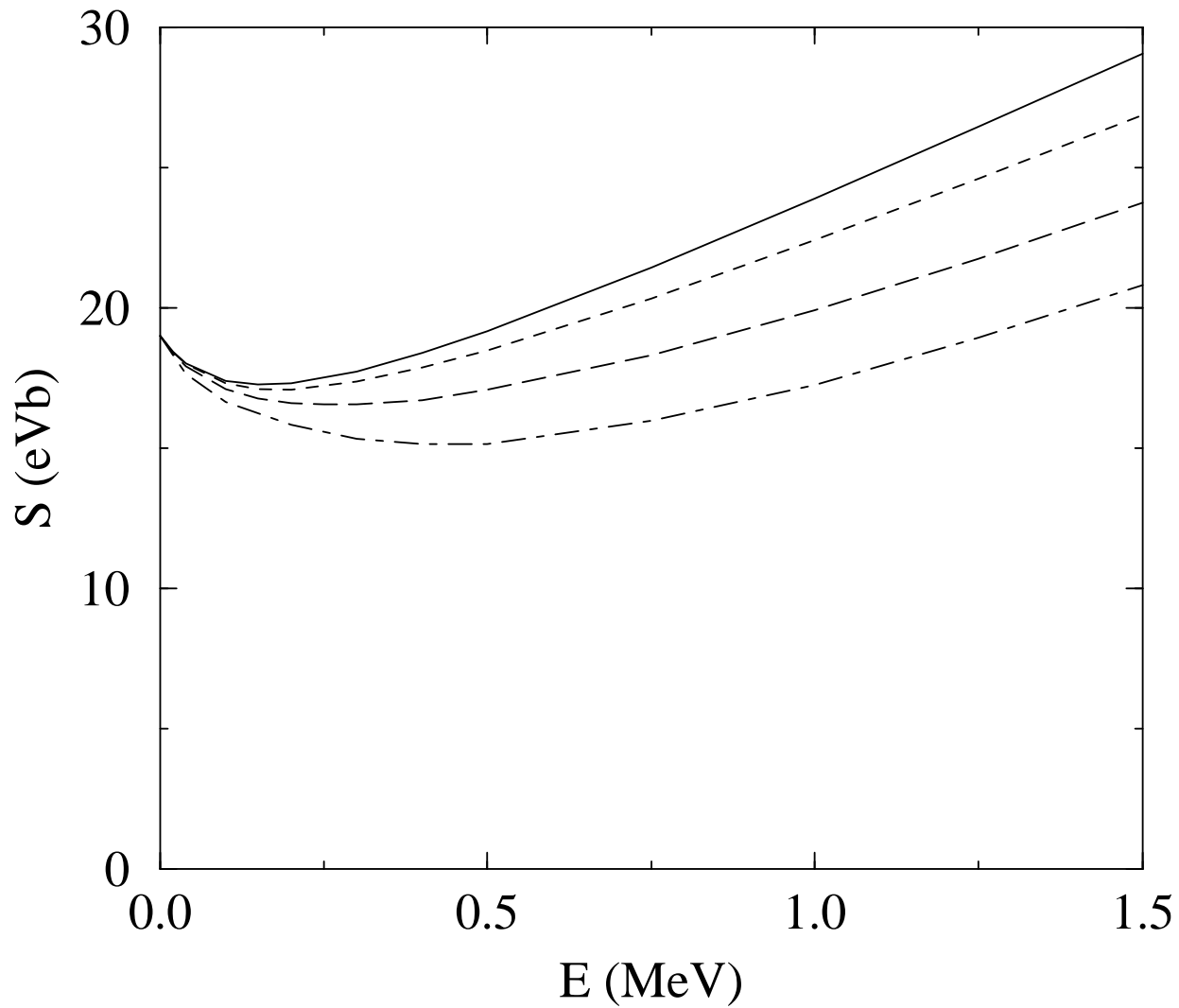,width=\linewidth}
\caption[]{The $S$ factor as a function of energy for a range of
cut-off radii. The results obtained for $r_c = 0.0$, 1.0, 2.4, and
4.1~fm are shown by the solid, short-dashed, long-dashed and
dot-dashed lines, respectively. All the curves are normalized to 19~eVb
at threshold.}
\label{fig-one}
\end{figure}

To further illustrate the role of nuclear distortions, the $S$ factors
for the cut-off radii of Table~\ref{table-padeb} are shown in
Fig.~\ref{fig-one}. The curves with $r_c = 0.0$, 1.0, 2.4, and 4.1~fm
are displayed by the solid, short-dashed, long-dashed, and dot-dashed
lines respectively. All the curves are normalized to 19.0~eVb at
threshold. The effect of nuclear distortion is quite noticeable even
at energies as low as 100~keV, and increases with increasing
energy. Fig.~\ref{fig-one}, together with Fig.~\ref{fig-two},
highlights an important aspect of the models: the equivalent hard
sphere radius is sensitive to the choice of potential
depth. Analogously, the phase shift and degree of nuclear distortion
are model-dependent. The effect of nuclear distortion is also seen in
Table~\ref{table-one} where the ratios $S(0)/S(20)$, $S(20)/S(100)$,
and $S(0)/A_n$ are given. The variation in the last ratio is almost
3\% indicating that the $S$ factor is sensitive to nuclear distortions
even at threshold.

Nunes {\em et al.} \cite{nunes} have also calculated the $S$ factor
with a Woods-Saxon potential, but used the Kim parameterization
\cite{kim}. We find good agreement with their calculations and, in
particular, concur with their observation of large effects due to
nuclear distortions in that particular model. The Kim parameterization
generates an $S$ factor with a slightly different energy dependence,
corresponding to a hard sphere model with $r_c = 3.0$~fm.

\begin{figure}
\centering\epsfig{file=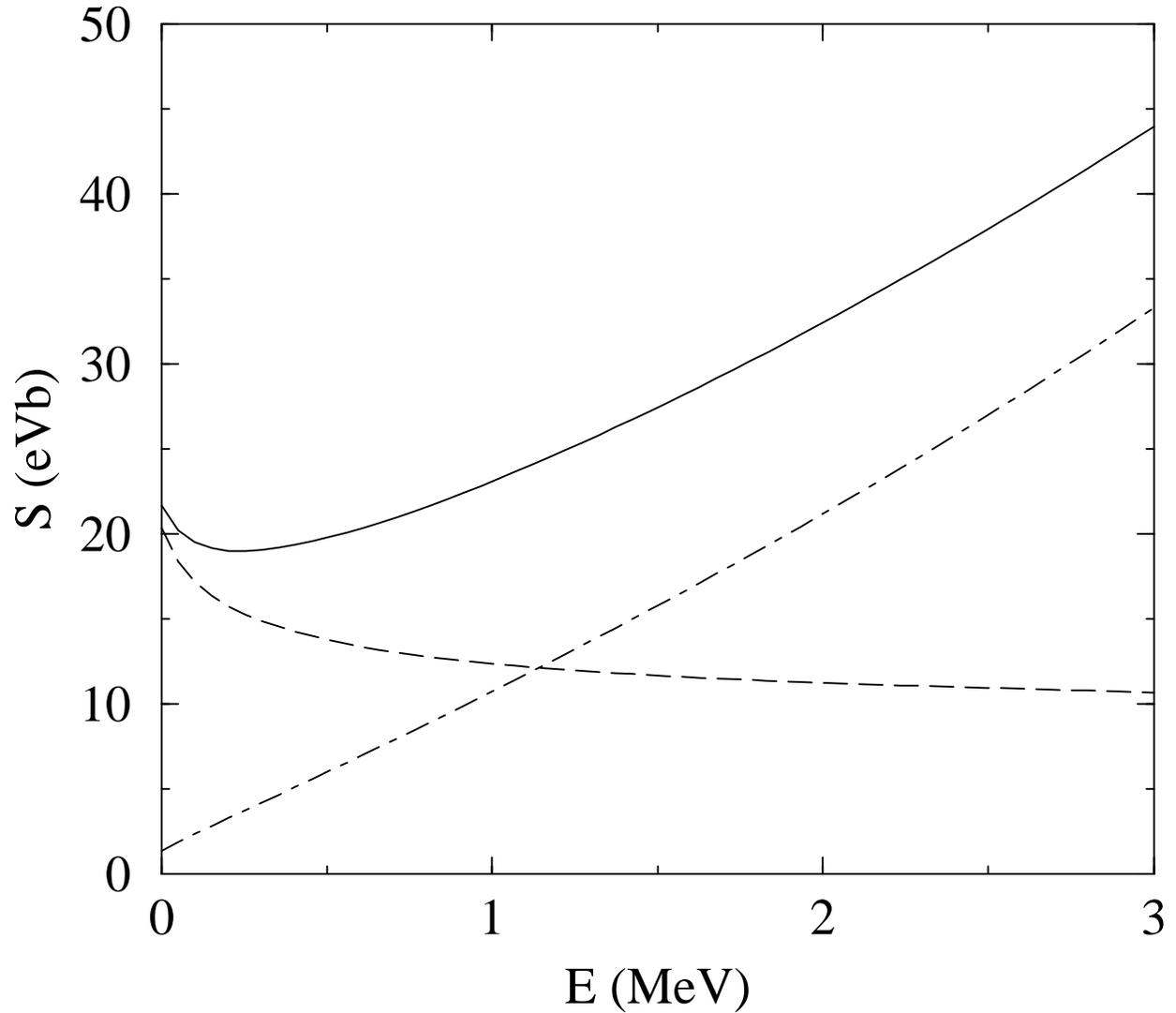,width=\linewidth}
\caption[]{The $S$ factor, as calculated with $r_c = 2.4$~fm. The
total result is displayed by the solid line, while the $s$- and
$d$-wave components are shown by the dashed and dot-dashed lines,
respectively.}
\label{fig-sd}
\end{figure}

The relative $s$- and $d$-wave contributions to the $S$ factor,
calculated for $r_c = 2.4$~fm, are displayed in Fig.~\ref{fig-sd}. The
total, $s$-wave, and $d$-wave parts are displayed by the solid,
dashed, and dot-dashed lines, respectively. The upturn at threshold is
purely from the $s$-wave component, even though the $s$- and $d$-wave
capture lead to the same final state. The linear behavior in the
$d$-wave component is a result of the zero in the Coulomb function
which lies very close to the position of the pole. In general, partial
waves for non-zero orbital angular momentum will have zeros on the
negative energy axis. The higher the angular momentum the closer they
will lie to threshold. Thus we do not expect to see an upturn when the
capture occurs from a high angular momentum state.

\section{Comparison with cluster models}
\label{sec-models}

The other class of model which has been used in the analysis of the
data are the cluster models
\cite{Calvin,desc,csoto-o,csoto,desca}. These generator-coordinate
models (GCM) calculate the $S$ factor microscopically, and incorporate
many-body effects which are not included explicitly in the simpler
one-body potential models. They predict the absolute magnitude as well
as the energy dependence. However, the use of the more sophisticated
models comes at a price: it is more difficult to discern the dominant
physical effects and to understand the differences between the various
calculations. Fortunately, the hard-sphere model can be used to
clarify these issues.

\begin{figure}
\centering\epsfig{file=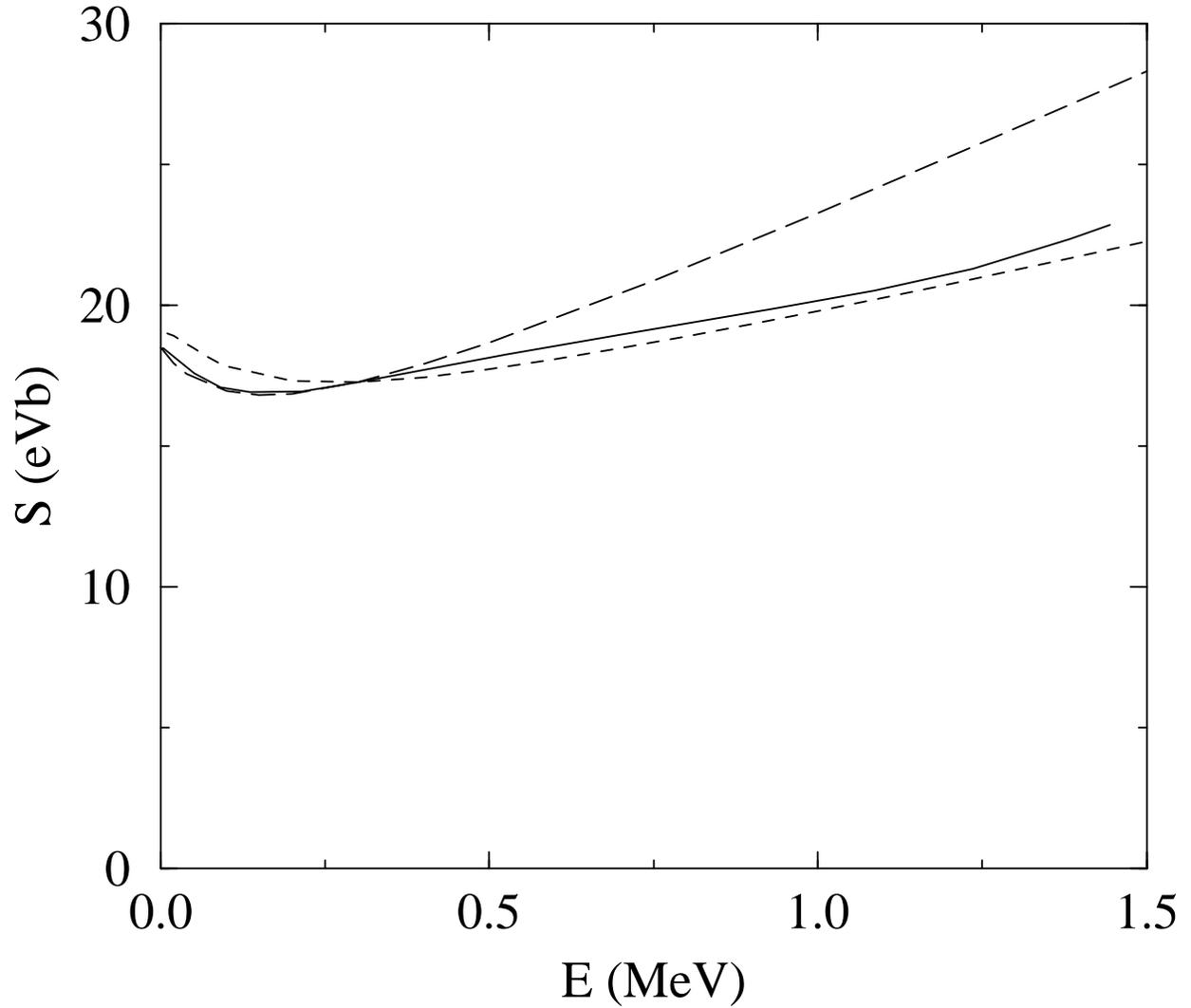,width=\linewidth}
\caption[]{The $S$ factor as a function of energy for various models.
The solid curve is result of the DB GCM calculation (as quoted by
Hammache {\em et al.}\protect \cite{Hammache}). The short-dashed curve
is the result of the C2B model calculation, while the long-dashed
curve is the hard-sphere model result with $r_c = 0$~fm. The curves
are normalized to agree at 0.3~MeV.}
\label{fig-desc}
\end{figure}

\begin{figure}
\centering\epsfig{file=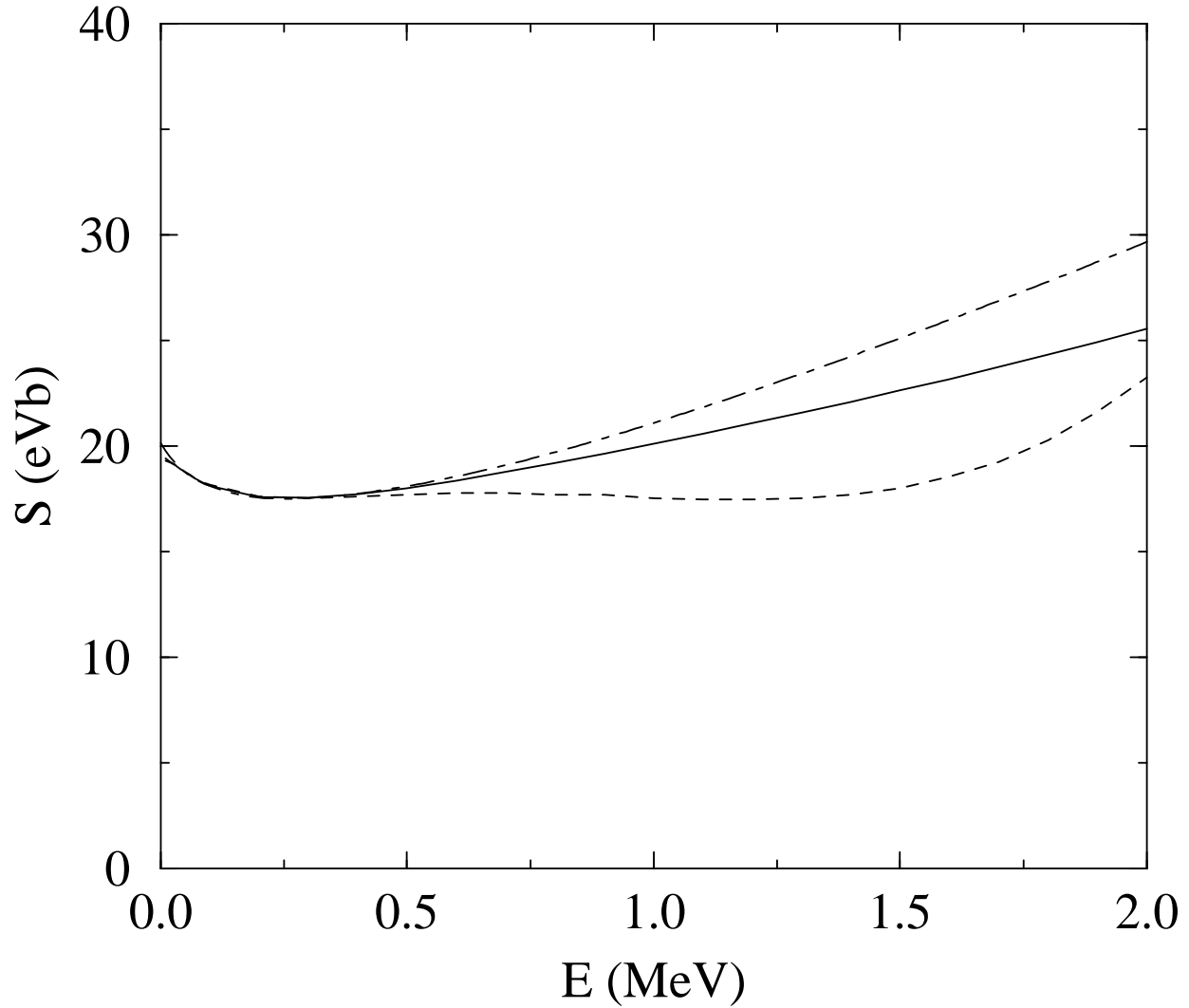,width=\linewidth}
\caption[]{The $S$ factor as a function of energy for various
models.  The C2B and C8B GCM model results are displayed by the solid
and dashed lines, respectively. The result of the $r_c=2.4$~fm
hard-sphere calculation is shown by the dot-dashed line.}
\label{fig-csoto}
\end{figure}

We compare the results of the hard sphere model calculation, with $r_c
= 0$~fm, with the GCM calculations of Descouvemont and Baye (denoted
as DB)\cite{desc} and the GCM calculations of Cs\'ot\'o {\em et al.}
\cite{csoto-o} (denoted as C2B) in Fig.~\ref{fig-desc}. The hard
sphere, DB, and C2B results are displayed by the long-dashed, solid,
and short-dashed lines, respectively. In the GCM calculation, C2B, the
effect of inter-cluster antisymmetrization, leading to effective
8-body wave functions, has not been included. All results have been
normalized to agree at 300~keV. Above this energy, the two GCM results
agree with each other, while below, the result of DB is consistent
with that of the $r_c = 0$~fm hard sphere model. However, the C2B
calculation is consistent with the hard sphere model result with $r_c
= 2.4$~fm, as shown in Fig.~\ref{fig-csoto}.  There, that result is
displayed by the solid line while the hard sphere model result is
displayed by the dot-dashed line. We show by the dotted line the
calculation, also by Cs\'ot\'o, which includes antisymmetrization
\cite{csoto}. This calculation will hereafter be referred to as the
C8B model. While this calculation shows the same energy dependence as
the C2B result near threshold, there is a marked change above 500~keV.

The differences between the C2B and C8B GCM calculations, manifest at
higher energies, have their source in both the $s$- and $d$-wave
contributions \cite{csoto-p}. The difference in the $s$-wave
calculations is relatively small and consistent with the expected
model dependence due to the different short range behavior. The effect
is much more dramatic for the $d$-waves; the contribution from three
of the $d$-wave channels goes to zero \cite{csoto-p} at approximately
1.5~MeV. That behavior is inconsistent with the very small effects of
nuclear distortion in the $d$-wave component observed in the one-body
models.

Outside the range of the nuclear force the wave functions, and hence
the matrix element, are determined by the properties of the Coulomb
force, the asymptotic strength, and the phase shift. As the two
calculations of Cs\'ot\'o have the same phase shift
\cite{csoto,csoto-p} the only difference must be at short distances,
less than $\sim 3$~fm. At 1.5~MeV, the integrand peaks at 14.5~fm, as
determined by the Coulomb properties and the phase shift. Therefore
the antisymmetrization, being the only difference between the two- and
eight-body models, must make the integrand very large at small
radii. That is possible if there is a resonance and, in that case, the
whole $d$-wave contribution may indeed vanish. This may be simulated
in the Woods-Saxon model by increasing the depth of the
potential. However, a resonance would have a very pronounced effect on
the phase shift, and there is no indication of a resonance in the
$d$-wave component. It is also inconsistent with the statement that
the two calculations of Cs\'ot\'o have the same phase shift.

We note that in the potential models antisymmetrization is not a
problem. One can construct a fully anti-symmetric eight-body wave
function for the $^7$Be-- $p$ system by taking a Slater determinant of
the bound states wave functions and the scattering wave
function. Orthogonality is assured between the wave functions when
they are calculated with the same potential, and the Slater
determinant is trivially constructed. As the transition operator is
one-body in our model only the incoming proton is involved in the
interaction with the mean field defined by the $^7$Be nucleus as a
whole. Antisymmetrization then becomes a problem only when one
ventures beyond the potential model and has wave functions that are
not orthogonal by construction.

\section{Comparisons with the data}
\label{sec-data}

The discussions in the previous sections have concentrated on
theoretical aspects of the low energy behavior of $S$ factor. We now
turn to the data to determine the normalization and a meaningful value
of the $S$ factor at astrophysical energies. When discussing the
$^7$Be($p,\gamma$)$^8$B $S$-factor data, it is important to consider
resonant capture. The main resonance in the reaction at low energies
is the $M1$ resonance at 0.637~MeV for the capture to the 0.774~MeV
state in $^8$B \cite{Aj88}. There is another, much weaker and wider,
resonance in the cross section at 2.183~MeV corresponding to the
2.32~MeV state in $^8$B \cite{Aj88}. We limit our comparisons to $E <
1.5$~MeV, to avoid the influence of this wider state, and assume the
resonance at 0.637~MeV to be of Breit-Wigner form.

There have been seven measurements of the $^7$Be($p,\gamma$)$^8$B $S$
factor. Two of these\cite{kav60,wiez} have large error bars and points
at only a limited number of energies, two and one respectively. This
limits their usefulness and we will not consider them further. The
remaining five measurements fall into two basic classes. The first is
comprised of the data of Kavanagh {\em et al.}  \cite{kavanagh} and of
Parker \cite{Parker}. Those data are on average $\sim 30$\% greater in
magnitude than the data of Filippone {\em et al.} \cite{Filippone},
Vaughn {\em et al.}  \cite{Vaughn}, and the most recent measurement of
Hammache {\em et al.} \cite{Hammache}. The areal density of the $^7$Be
target is frequently determined by measuring the $^7$Li($d,p$) cross
section, $\sigma_{dp}$, and normalizing the data to the known
value. We follow the recommendation of \cite{adelberger} and take
$\sigma_{dp} = 147 \pm 11$~mb for all the data sets. Strieder {\em et
al.}  \cite{strieder} quote smaller errors, $\sigma_{dp} = 146 \pm 5$,
while a recent measurement \cite{Weissman} gives $\sigma_{dp} = 155
\pm 8$~mb. The use of the different values for this normalizing
reaction shifts the results only slightly. The measurements of
Filippone {\em et al.} \cite{Filippone} and of Hammache {\em et al.}
\cite{Hammache} also determined the density by direct measure of the
$\beta$-delayed $\gamma$ rays, and found very good agreement between
the two normalization methods.

To extract a reliable value of the $S$ factor at astrophysical
energies we must carefully consider the experimental errors.  Some of
the errors are common between different data sets: for example the
value of $\sigma_{dp}$ used in the normalization. Also in fitting
individual data sets errors common to all the points must be handled
separately. Thus we fit each data set separately using just the
relative errors. The sets used are Kavanagh {\em et al.}
\cite{kavanagh}, Vaughn {\em et al.} \cite{Vaughn}, Filippone {\em et
al.} \cite{Filippone}, and the two data sets of Hammache {\em et al.}
\cite{Hammache}. That measurement reports two data sets from runs
taken in different years with different normalization errors. Hence,
the two Hammache data sets are fit separately using just the relative
errors and combined using both relative and absolute errors.  In
averaging the values of $S(20)$ obtained from the fits to the Vaughn
and Filippone data we have taken into account the common error due to
the uncertainty in $\sigma_{dp}$ used for the normalization.

The resonance at 0.637~MeV was fit with the Breit-Wigner form.  In
comparisons to the data of Filippone {\em et al.} \cite{Filippone} and
Kavanagh {\em et al.}  \cite{kavanagh} the parameters of the
Breit-Wigner were determined by the fit to the data. In fitting the
data of Vaughn {\em et al.}  \cite{Vaughn} and Hammache {\em et al.}
\cite{Hammache} the resonance parameters were taken from the fit to
the Filippone data. Leaving out the resonance in the fit of the
Hammache data raises the result by $2-3$\%.

\begin{figure}
\centering\epsfig{file=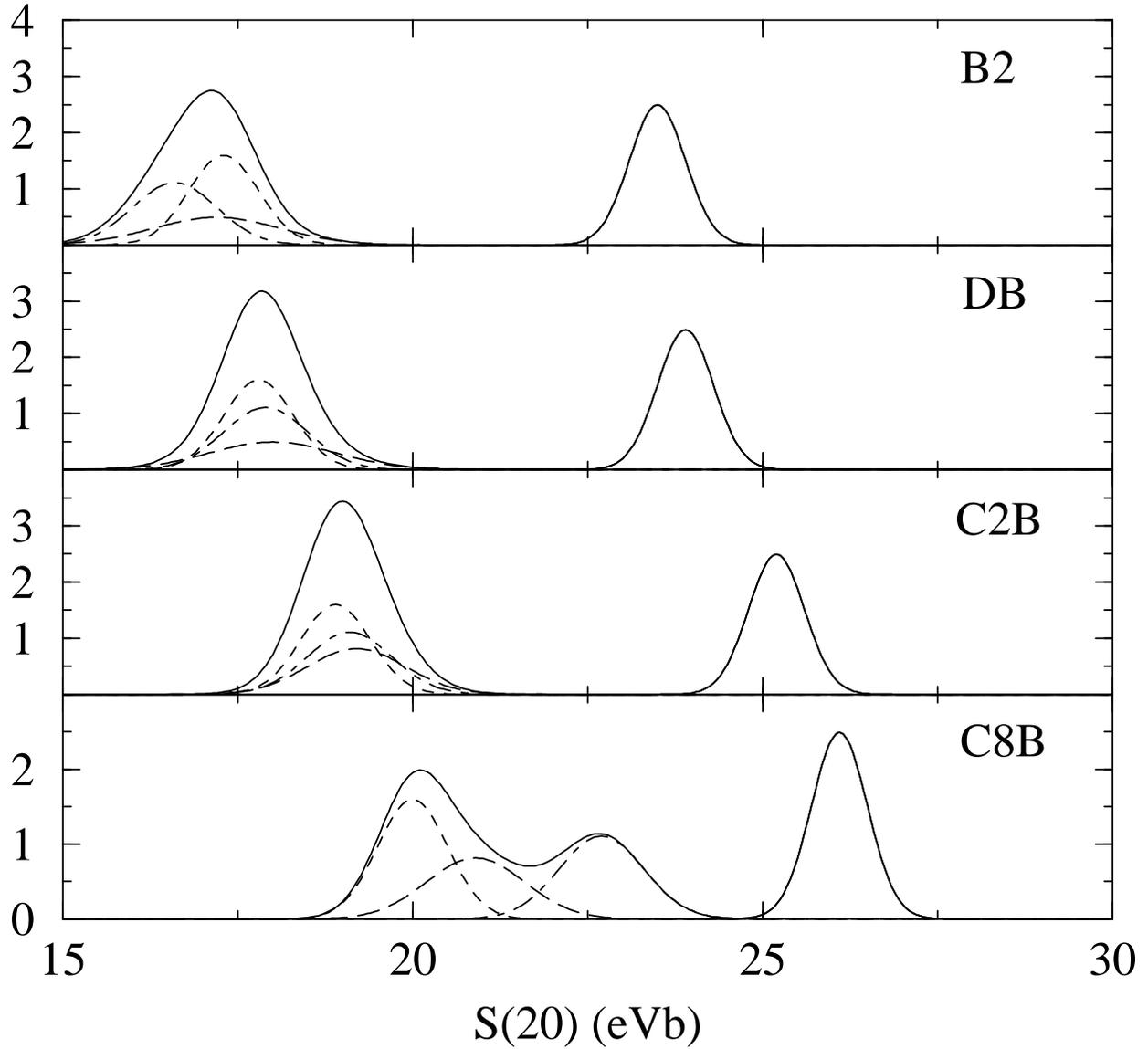,width=\linewidth}
\caption[]{Ideogram\protect\cite{pdg} for the $S$ factor at 20~keV.
The calculations are denoted by B2, DB, C2B, and C8B as defined in the
text. The results obtained from the data of Filippone {\em et al.},
Vaughn {\em et al.}, and Hammache {\em et al.} are displayed by the
short-dashed, dot-dashed, and long-dashed lines respectively. The
solid line is the sum of the contributions. The isolated peak at
higher $S$ values is from Kavanagh {\em et al.} \cite{kavanagh}}
\label{fig-ideo}
\end{figure}

The values of $S(20)$ determined as a result of the fits to the data
are shown in Table~\ref{table-S} for hard sphere ($r_c = 2.4$ and
1.0~fm), Woods-Saxon (B1 and B2) and GCM (DB, C2B, and C8B)
models. For each model two fits are shown: the left column for each
data set is the result obtained by fitting the data below 400~keV,
while the right column is the result obtained by fitting up to
1.5~MeV. As the data of Vaughn {\em et al.} do not extend below
400~keV only the result of fitting to the higher energy is presented
in that case. The final column lists the result of the average of the
Vaughn {\em et al.}, Filippone {\em et al.}, and Hammache {\em et al.}
fits. The ideograms \cite{pdg} for the B2, DB, C2B, and C8B results
are displayed in Fig.~\ref{fig-ideo} from top to bottom in the order
listed. Therein, the results obtained from fitting the data of
Filippone {\em et al.}, of Vaughn {\em et al.}, and of Hammache {\em
et al.}  below 1.5~MeV are displayed by the short-dashed, dot-dashed,
and long-dashed lines respectively. The solid line is the sum of all
four Gaussian distributions; the peak at $\sim 25$~eVb was obtained
from the Kavanagh data set. As indicated by the disparity in the
absolute magnitude, the results obtained from the Kavanagh data are
consistently much higher than those obtained from the other data
sets. Hence it was not included in the average given in
Table~\ref{table-S}.

\begin{figure}
\centering\epsfig{file=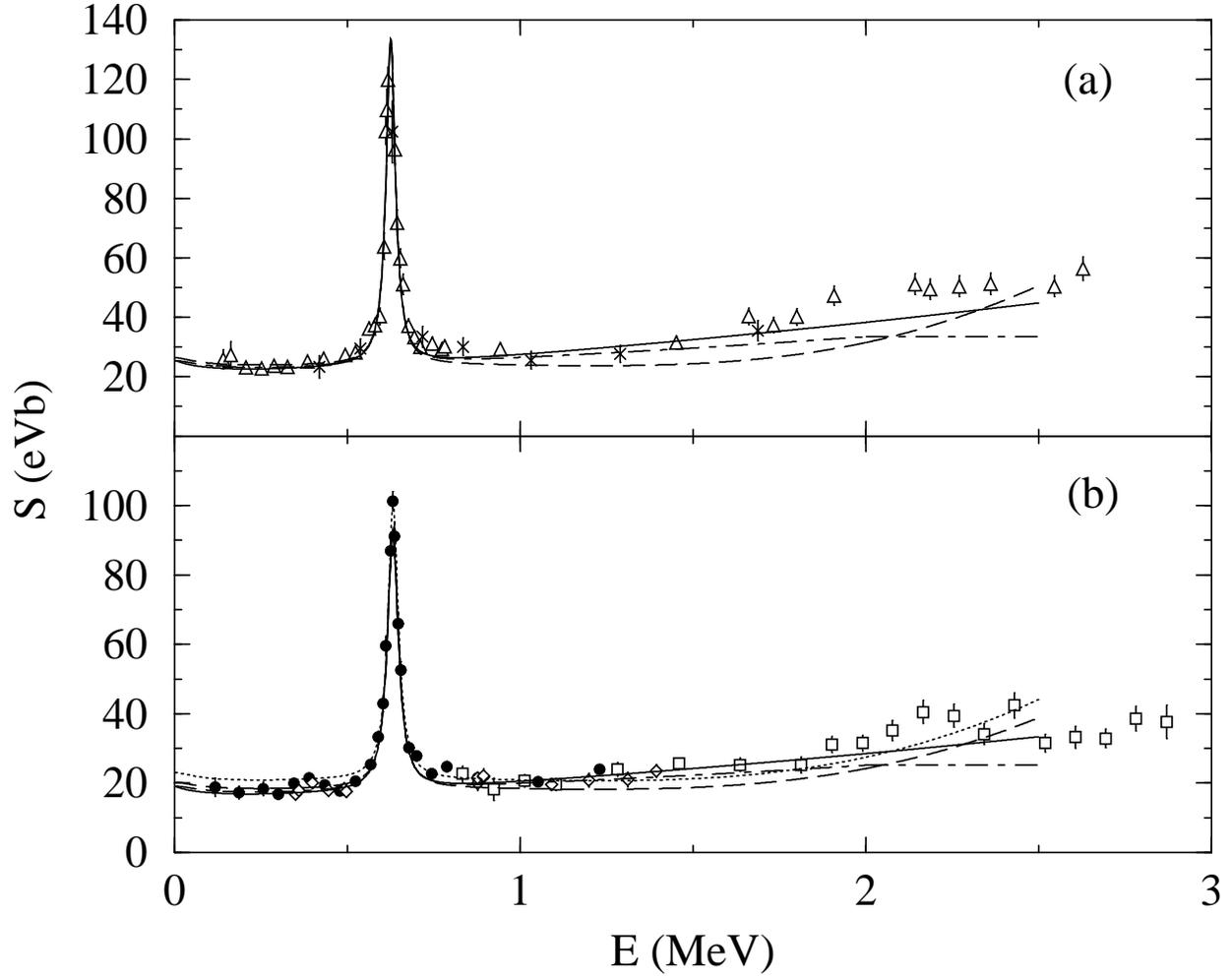,width=\linewidth}
\caption[]{The theoretical $S$ factor as function of energy compared to
the experimental data. In (a) we present the data of Kavanagh {\em et
al.} \protect\cite{kavanagh} while in (b) we show the data of Vaughn
{\em et al.} \protect\cite{Vaughn}, Filippone {\em et al.}
\protect\cite{Filippone} and Hammache {\em et al.}
\protect\cite{Hammache}. In each figure, the solid line shows the B1
model result, while the dot-dashed and dashed curves are the results
of the C2B and C8B calculations, respectively. In (b) the fits are to
the Filippone data, with the exception of the dotted curve in (b)
which is a fit to the Vaughn data using the C8B calculation.  All the
fits are to the 0--1.5~MeV energy range.}
\label{fig-fitss}
\end{figure}

\begin{figure}
\centering\epsfig{file=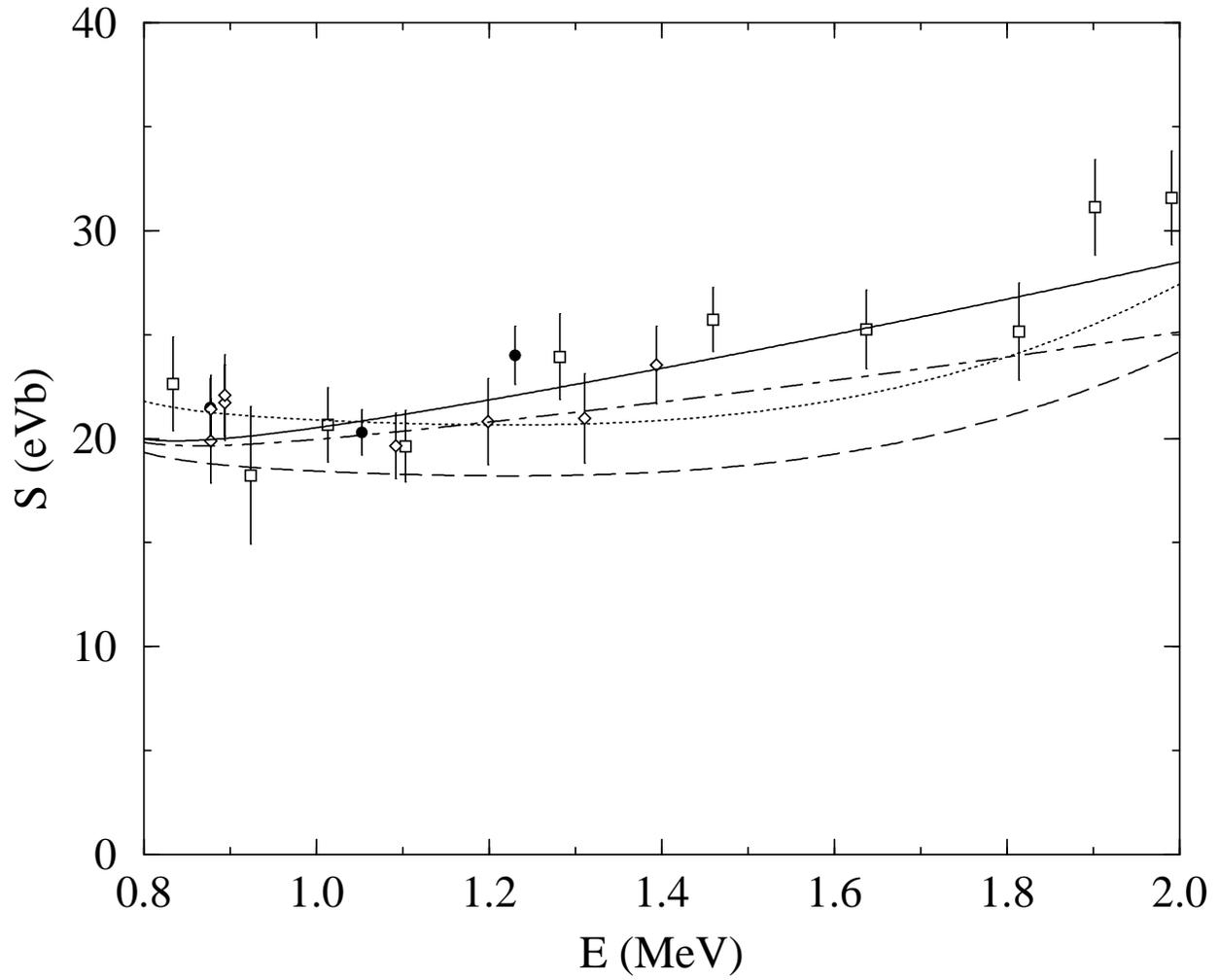,width=\linewidth}
\caption[]{Expanded view of the $S$ factor as function of energy.
This is an expanded view of Fig.~\ref{fig-fitss}(b) using the same
conventions for the curves and data points.}
\label{fig-zoom}
\end{figure}

The best consistency in the values of $S(20)$ among the data sets,
excluding the Kavanagh data, is found with the fits using the C2B and
DB models. The high quality of the fit to the data using the B1 and
C2B models can be seen in Figs.~\ref{fig-fitss} and
\ref{fig-zoom}. The quality of fit using the DB model is similar and
is not shown. Note that the values obtained from the DB calculation
are consistently lower than the C2B values by $3-6$\%. Our DB result
agrees with that quoted by Hammache {\em et al.} \cite{Hammache} who
also used this calculation in their extrapolation. The GCM model
calculation of Johnson {\em et al.}  \cite{Calvin} has a very similar
energy dependence to the DB calculation and would give similar results
in a fit. The low values of $S(20)$ obtained with these models,
compared to the C2B and C8B models, is a consequence of the energy
dependence near threshold. As was shown in Fig.~\ref{fig-desc}, the DB
model calculation agrees with the hard sphere model for $r_c = 0$~fm
in the low-energy region.  The B2 Woods-Saxon model fit looks equally
good but the extrapolated value of the $S$ factor depends on the
energy range used in the fit. The value drops by 5\% for the larger
energy range.

The values of $S(20)$ obtained from fitting the C8B model to the data
sets show much more dispersion than any of the other results, varying
with both the energy range and data set used. The latter variation is
clearly seen in the C8B ideogram, Fig.~\ref{fig-ideo}. This is
reflected in Figs.~\ref{fig-fitss} and \ref{fig-zoom}, where the C8B
calculation has the wrong energy dependence when compared to {\em any
given data set}. While the data suggests a steady increase in the $S$
factor above 1~MeV, the C8B result is relatively flat in the 1 to 2
MeV range then shows a sharp increase above 2~MeV. This disagreement
with the data must be understood before using this model to
extrapolate the data to threshold.

In Table~\ref{table-chi} we show the $\chi^2$ per degree of freedom
for individual data sets for four different models. Only data below
1.5~MeV has been included in the fit. Each data set has a common
normalization error which is not included in the calculation of
$\chi^2$ . For the data of Hammache {\em et al.} \cite{Hammache} we
fit only the 1996 data. The higher value of $\chi^2$ per degree of
freedom for the data of Kavanagh {\em et al.} \cite{kavanagh} is due
to three points on the lower side of the resonance peak; the fitting
ignores any relative error in the energy calibration. For the other
three data sets the $\chi^2$ per degree of freedom is of order one
except for the C8B results for which it is consistently
higher. This reflects the incorrect energy dependence of the C8B
model noted in the previous paragraph.

\begin{figure}
\centering\epsfig{file=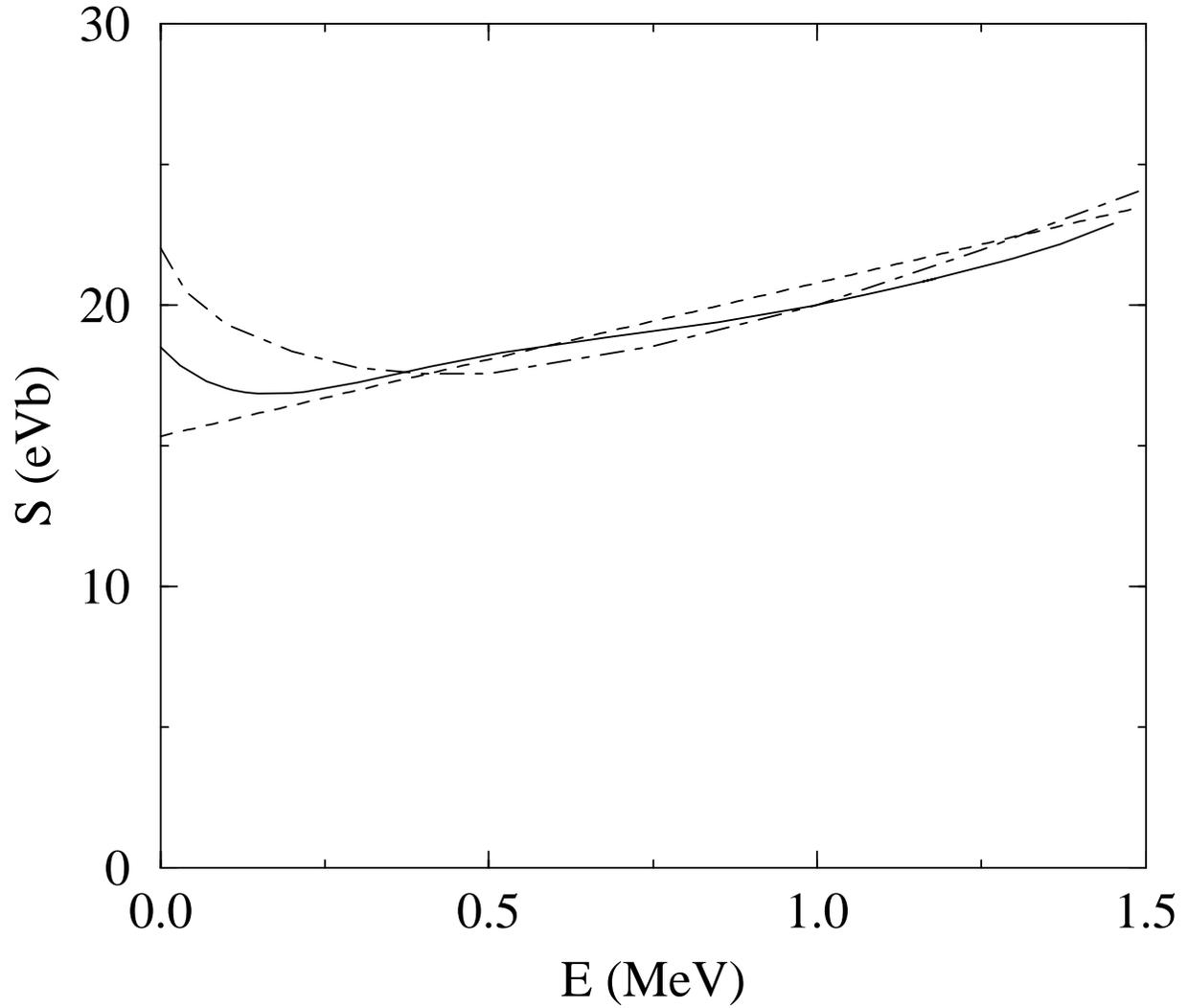,width=\linewidth}
\caption[]{The $S$ factor as function of energy. The result of the
DB calculation, the straight line fit from Fig.~\ref{fig-fits}, and the $r_c =
4.1$~fm model calculation are displayed by the solid, dashed, and
dot-dashed lines respectively.}
\label{fig-csoto-sc}
\end{figure}

We may now discuss in more detail Fig.~\ref{fig-fits}, where two
extreme fits to the non-resonant part of the data are shown. The first
is the naive straight-line fit while the other is the hard sphere
model result with $r_c = 4.1$~fm. These two results are compared to
the DB GCM calculation in Fig~\ref{fig-csoto-sc}. The resonant
contribution has been removed from all the curves. All three results
are in reasonable agreement with each other in most of the energy
region where data exists, $E>0.1$~MeV; hence the equally good
fits. However, only the DB calculation contains the correct physics at
low energy. The straight line fit does not contain any contribution
from the pole term and hence no upturn, while the upturn in the $r_c =
4.1$~fm hard sphere model calculation is far too severe. Also, the
curvature in that hard sphere model result is not seen in the GCM
model calculations in the region of the data.

We can improve on the straight line fit by replacing the constant term
with $\left\{ 0.0408/(0.1375 + E) + 0.7033 \right\}$ from the
rational approximation, Eq.~(\ref{eq-pade}). Since the ratio of the
pole to constant term is fixed in an almost model independent manner
and the pole term, by itself, is poorly determined by the data it make
little sense to fit these two terms separately.  This leads to the form
\begin{equation}
S(E) = S(0) \left[ \left( \frac{0.0408}{0.1375 + E} + 0.7033 \right) +
c'E \right]
\label{eq-pade-f}
\end{equation}
where $S(0)$ and $c'$ are fit parameters and $E$ is in MeV. Using this
form and a Breit-Wigner resonance we fit the data up to 1.5~MeV. The
resulting fit [$S(0) = 18.5$~eVb and $c' = 0.351$~MeV$^{-1}$] is as good as
that with the straight line but the value of $S(20)$ is now $18.0 \pm
0.9$~eVb. This is in agreement with the DB results when fit over the
same energy range. The parameter $c'$, determined by the fit, is also
close to that obtained from the B2 potential model and corresponds to
a low value of $r_c$ (see Table~\ref{table-padeb}).

\section{Conclusions}
\label{sec-conclusions}

We have explored the energy dependence of the $^7$Be($p,\gamma$)$^8$B
reaction and have constructed a simple model to illustrate the
dominant physics. For energies, $E < 0.4$~MeV, the $S$ factor is
dominated by a pole arising from the influence of the sub-threshold
$^8$B ground state. The behavior of the $S$ factor near threshold is
determined by three parameters: the location and residue of the pole,
and by the effective hard sphere radius. The location of the pole is
set by the binding energy of the valence proton in $^8$B, while the
residue is proportional to the normalization of the $S$ factor. The
hard sphere radius simulates the effect of the nuclear distortion, and
changes the slope of the $S$ factor. The low energy behavior of the
models considered in this paper can be reproduced with $r_c$ in the
range 0 to 3~fm. If we restrict the comparison to models that fit the
elastic scattering of neutrons from $^7$Li a value of 1~fm is
preferred. The $S$-factor data do not distinguish between the results
obtained with these values. While some models at low energy are
consistent with a hard sphere model of $r_c = 0$~fm, suggesting no
nuclear distortion at threshold, this is not of major concern: the
preferred value of 1~fm is suitably well inside the nucleus so that
the influence of nuclear distortion for what is, in effect, a surface
capture is minimal. The range of $r_c$ from 0 to 3~fm introduces an
error of about 5\% in the extrapolations even when only the data below
400~keV is used. Restricting ourselves to models consistent with the
elastic neutron $^7$Li scattering data reduces this uncertainty to
about 1\%.  Thus from fitting the energy range of 0 to 400~keV we have
$S(20) = 18.4\pm 1.0\pm0.2$~eVb, or equivalently $S(0) = 19.0\pm
1.0\pm0.2$~eVb, where the first error is experimental and the second
is from model dependences in the fit.

As the energy increases beyond 400~keV the model dependencies and,
hence, uncertainties also increase. Yet even at the somewhat higher
energy region $E < 2$~MeV the simple model gives indications of how
large the model effects might be. For example, the effects of nuclear
distortion in the $d$-wave component of the $S$ factor is at most 2\%
even for $E \sim 1.5$~MeV, as the scattering wave barely penetrates
into the region where the nuclear forces are strong. Together with the
comparison with existing data, this casts doubt on the eight-body
model result of Cs\'ot\'o.

Of the models considered here the DB and C2B GCM calculations have the
least variation of the extracted $S(20)$ with the energy range fit,
suggesting they do have a more accurate description of the physics in
the higher energy range.  However the GCM calculations predict the
magnitude of $S(20)$ to be much higher than that we extract from the recent
experimental data. As with the fit to the lower energy range we
restrict ourselves to models that are consistent with the neutron
$^7$Li elastic scattering data.  Elastic $p$-$^7$Be scattering data
would be useful in confirming the conclusions drawn from the elastic
$n$-$^7$Li scattering data.

Neither data nor theoretical considerations are sufficiently refined
to completely rule out the Woods-Saxon models. The value of $S(20)$
differs by 5\% depending on whether the GCM or the Woods-Saxon model
is used for the extrapolation.  Taking this as an indication of the
error for the fit in the energy range $E < 1.5$~MeV we then obtain
$S(20) = 17.6 \pm 0.7 \pm 0.4$~eVb and $S(0) = 18.1 \pm 0.7 \pm
0.4$~eVb. The second error comes from the uncertainty (half the
spread) in the choice of model used in the extrapolation. As the
theoretical error here is not well understood we prefer the values of
the $S$ factor extracted from the more restricted energy range. Phase
shift information from proton $^7$Be scattering in the 1~MeV region
would reduce the theoretical uncertainty from the fit over the larger
energy range.

\acknowledgements{The Natural Sciences and Engineering Research
Council of Canada is thanked for financial support. We thank
G.~Bogaert for supplying the experimental data from
Ref.~\cite{Hammache}.  E.~Adelberger, D.~Baye, L.~Buchmann,
A.~Cs\'ot\'o, C.W.~Johnson, G.A.~Miller, K.~Snover and E.W.~Vogt are
thanked for useful discussions. We also thank D.~Baye for suggesting
improvements to the manuscript.}

\begin{table}
\caption[]{The effects of nuclear distortion on the direct capture matrix
elements using a Woods-Saxon potential. The calculations are as
discussed in the text.}
\label{table-3}
\begin{tabular}{ccccc}
model              &  \multicolumn{2}{c}{$s$-wave}  &
\multicolumn{2}{c}{$d$-wave} \\
Energy             & 0.5~MeV & 1.5~MeV & 0.5~MeV & 1.5~MeV \\
\hline
Full distortion    & 517.5   & 115.6   & 241.1   & 95.12   \\
Partial distortion & 525.2   & 121.8   & 242.2   & 97.59   \\
No distortion      & 550.7   & 135.1   & 240.7   & 93.58   \\
Pure Coulomb       & 560.4   & 144.5   & 240.8   & 93.77   
\end{tabular}
\end{table}

\begin{table}
\caption[]{The percentage error obtained by using various expansion of
the $S$ factor for $r_c = 2.4$~fm. The errors introduced by using a
rational approximation [Eq.~(\ref{eq-pade})], a second order Taylor
series expansion, a third order Taylor series expansion, and the
logarithmic expansion of Ref.~\cite{williams} are denoted by $\delta
S_r$, $\delta S_{T2}$, $\delta S_{T3}$, and $\delta S_W$ respectively.}
\label{table-pade}
\begin{tabular}{crrrr}
$E$ (MeV) & $\delta S_r$ (\%) & $\delta S_{T2}$ (\%) & 
$\delta S_{T3}$ (\%) 
& $ \delta S_W $ (\%) \\ 
\hline
0.00 &   0.00 &    0.00 &     0.00 &    0.00\\
0.02 &  $-$0.02 &    0.06 &    $-$0.03 &    0.04\\
0.10 &  $-$0.11 &    7.22 &    $-$5.03 &    5.34\\
0.30 &  $-$0.36 &  110.78 &  $-$230.92 &  124.51\\
0.40 &  $-$0.56 &  211.84 &  $-$590.43 &  384.07\\
0.50 &  $-$1.40 &  340.70 & $-$1191.88 & 1259.63\\
1.00 &  $-$6.64 & 1309.96 & $-$9214.33 & $1.4\times10^7$
\end{tabular}
\end{table}

\begin{table}
\caption[]{Coefficients of the expansion Eq.~(\ref{eq-pade}).  Also
shown are the first and second logarithmic derivatives at
threshold. The first four rows are for the hard sphere model with
different cut-off radii. The fifth row contains the results from
Williams and Koonin \protect\cite{williams} while the sixth, seventh and
eight rows are with Woods-Saxon potentials. The last two rows
contain values of the derivatives from previous work.}
\label{table-padeb}
\begin{tabular}{cccccc}
Model & a (MeV) & b & c (MeV$^{-1}$) & $d\log{S}/dE$ (MeV$^{-1}$) & $
d^2\log{S}/dE^2$ 
(MeV$^{-2}$) \\
\hline 
$r_c = 0.0$~fm & 0.0409 & 0.702 & 0.390 & $-$1.77 & 28.3 \\
$r_c = 1.0$~fm & 0.0409 & 0.703 & 0.343 & $-$1.82 & 28.1\\ 
$r_c = 2.4$~fm & 0.0408 & 0.703 & 0.239 & $-$1.92 & 27.7 \\ 
$r_c = 4.1$~fm & 0.0407 & 0.704 & 0.067 & $-$2.09 & 27.0 \\ 
W \& K \cite{williams}
               & 0.0425 & 0.687 & 0.050 & $-$2.35 & 28.3 \\ 
B1 \cite{barker1} 
               & 0.0420 & 0.695 & 0.310 & $-$1.91 & 28.6 \\
B2 \cite{barker1}
               & 0.0417 & 0.697 & 0.367 & $-$1.84 & 28.7 \\
T \cite{tombrello} 
               & 0.0409 & 0.703 & 0.200 & $-$1.96 & 27.6 \\
Barker \cite{barker2}
               &  ---   & ---   & ---   & $-$1.41 & 10.2 \\
Adelberger {\em et al.} \cite{adelberger}
               &  ---   & ---   & ---   & $-$0.70 & \phantom{0}3.3  \\   
\end{tabular}
\end{table}  

\begin{table}
\caption[]{The energy dependence of the $S$ factor at low energy for
various cut-off radii and Woods-Saxon models.}
\label{table-one}
\begin{tabular}{cccc}
Model & $S(0)/S(20)$ & $S(20)/S(100)$ & $S(0)/A_n$ \\
\hline
$r_c = 0.0$~fm     & 1.03 & 1.06 & 38.2 \\
$r_c = 1.0$~fm     & 1.03 & 1.06 & 38.1 \\
$r_c = 2.4$~fm     & 1.03 & 1.07 & 37.8 \\
%$r_c = 2.9$~fm     & 1.03 & 1.08 & 37.6 \\
$r_c = 4.1$~fm     & 1.04 & 1.10 & 37.2 \\
B1 \cite{barker1}  & 1.03 & 1.08 & 37.8 \\
B2 \cite{barker1}  & 1.03 & 1.06 & 38.0 \\
T \cite{tombrello} & 1.03 & 1.08 & 37.6
\end{tabular}
\end{table}

\begin{table}
\caption[]{Values of $S(20)$ in eVb extracted from the various experimental
data sets with different models. For each data set, except Vaughn, we
have two columns showing the extracted value of $S(20)$ using data up
to 400~keV and to 1.5~MeV respectively. The last two columns give the
weighted average of the $S$ factor obtained from the Vaughn, Filippone
and Hammache data sets. The last row is the standard deviation errors
on the value of the $S$ factor including both statistical and
normalization errors.}
\label{table-S}
\begin{tabular}{lccccccccc}
Model & \multicolumn{2}{c}{Kavanagh} & Vaughn &
\multicolumn{2}{c}{Filippone} & \multicolumn{2}{c}{Hammache} &
\multicolumn{2}{c}{Average}\\\hline 
 $r_c = 2.4$~fm     & 24.3& 25.1& 18.1& 19.3& 18.7& 19.5& 18.7& 19.4& 18.6\\ 
 B1       & 24.1& 24.7& 17.7& 19.1& 18.4& 19.3& 18.4& 19.2& 18.3\\
 C2B      & 24.0& 25.2& 19.1& 19.1& 18.9& 19.3& 19.2& 19.2& 19.1\\
 C8B      & 24.0& 26.1& 22.7& 19.2& 20.0& 19.4& 20.9& 19.3& 21.0\\
 $r_c = 1.0$~fm     & 23.3& 23.3& 16.0& 18.5& 17.1& 18.5& 16.9& 18.5& 16.9\\ 
 B2       & 23.3& 23.5& 16.6& 18.5& 17.3& 18.4& 17.2& 18.4& 17.2\\
 DB       & 23.2& 23.9& 17.9& 18.4& 17.8& 18.3& 18.0& 18.3& 18.0\\
$\sigma $ &  1.9&  1.9&  1.5&  1.6&  1.5& 1.4 & 0.9 &  1.1&  0.8 
\end{tabular}
\end{table}

\begin{table}
\caption{The $\chi^2$ per degree of freedom for fits to individual
data sets. Each fit is to a data set with common normalization errors
and only the relatives errors are included. Only data below 1.5~MeV
has been included in the fits. }
\label{table-chi}
\begin{tabular}{ccccc}
Model & Kavanagh & Vaughn & Filippone & Hammache 96 \\\hline
B1    & 2.5      &  0.6   &    1.0    &    0.8     \\
C2B   & 2.6      &  0.9   &    1.0    &    0.9     \\
C8B   & 3.7      &  1.6   &    1.8    &    2.7     \\
DB    & 2.4      &  0.7   &    1.0    &    0.8     
\end{tabular}
\end{table}

\end{document}